\documentclass[aps,prx,reprint,superscriptaddress,amsmath,amssymb,floatfix,longbibliography]{revtex4-2}

\usepackage{graphicx}
\usepackage{bm}
\usepackage{amsmath,amssymb}
\usepackage{xcolor}
\usepackage{hyperref}

\newcommand{\kt}{k_BT}

\begin{document}

\title{Kinetics, Proofreading, and the Lifetime of Multivalent Contacts}

\author{Giuseppe Battaglia}
\email{gbattaglia@ibecbarcelona.eu}
\affiliation{Catalan Institution for Research and Advanced Studies (ICREA), Barcelona, Spain}
\affiliation{Institute for Bioengineering of Catalonia (IBEC), The Barcelona Institute of Science and Technology (BIST), Barcelona, Spain}

\date{\today}

\begin{abstract}
Selective recognition must also be reversible: a cell must hold its target tightly enough to tell it apart, then let go to act again, and equilibrium binding, which weighs only how favorable a contact is, cannot do both. What settles recognition is how long a contact lasts. We make lifetime the control variable, from a master equation through Smoluchowski to the Kramers lifetime of a multivalent contact. Lifetimes are exponential in free energy, so a tenfold change in receptor number moves a contact from seconds to months, turning a thermodynamic gradient into a switch. Selectivity then depends on the encounter: under flow a particle is kept only if its contact outlives the vascular dwell time, which sharpens the density threshold. Equilibrium also caps discrimination, a ceiling cells break by spending energy, and what the drive buys depends on the valency of the contact it reads. On a single bond, as in T cell antigen discrimination, a driven cascade turns a small free-energy difference into orders of magnitude; immune memory lowers the activation threshold and leaves specificity untouched. On a multivalent contact the lifetime is itself exponential in receptor density, so the cascade raises that exponential to the number of driven steps: endocytosis proofreads what it takes up, a double amplification that vanishes without ATP and points to phenotype-selective delivery.
\end{abstract}

\maketitle

\section{Introduction}\label{sec:intro}

Equilibrium statistical mechanics has been remarkably successful at explaining multivalent recognition. The combinatorial entropy of binding, counted correctly, produces thresholds in receptor density that no monovalent ligand can match~\cite{Kitov2003,Martinez-Veracoechea2011}; those thresholds have been measured for polymers, colloids and DNA nanostructures~\cite{Dubacheva2023,Linne2021,Bila2022}, and they have been exploited to build carriers that read receptor density rather than receptor identity~\cite{Tian2020,Liu2020}. The equilibrium theory of this regime, two multivalent units negotiating in crowded water, has been worked out in detail elsewhere~\cite{PATarxiv}, building on the superselectivity literature~\cite{Martinez-Veracoechea2011,Dubacheva2023,Curk2025}. It yields a free-energy functional whose every term is separately measurable, and that functional is the starting point here. Section~\ref{sec:background} restates the parts of it we need.

But a free energy answers exactly one question: given unlimited time, what fraction of encounters end up bound? For a great deal of chemistry that is the right question. For biology, and for nanomedicine, it very often is not, and there are three independent reasons.

The first is that recognition does not have unlimited time. A nanoparticle in the bloodstream passes a given endothelial surface in a fraction of a second. A T cell scans a dendritic cell for minutes, examining thousands of peptides. A virion has one attempt before mucociliary clearance removes it. If the encounter is shorter than the time needed to reach equilibrium, then what matters is not the depth of the free-energy minimum but the rate of descent into it. Two constructs with identical equilibrium constants can behave completely differently under flow, and the one that wins is the one that engages faster.

The second reason is that a deep minimum is not a bound state if the system cannot get into it, and not a useful bound state if it cannot get out. Two bare surfaces in physiological salt fall into a dispersion well some $21\,\kt$ deep~\cite{PATarxiv}. Thermodynamically that is spectacular binding. Functionally it is a disaster, because nothing that binds that hard can ever be released, recycled or regulated. Living systems, and well-designed carriers, need contacts that last a specified time, and the specified time is a kinetic quantity that the equilibrium constant does not determine.

The third reason is the deepest. Some discriminations are provably impossible at equilibrium and are achieved only by spending energy. The classic example, which we take up in Sec.~\ref{sec:tcell}, is T cell antigen recognition, where discrimination between a pathogenic peptide and a self peptide differing by a single residue far exceeds what their binding free energies would permit. The trick is to make the decision depend on how long a complex survives rather than on how favorable it is, and to read that lifetime out through a sequence of energy-consuming steps. This is kinetic proofreading~\cite{Hopfield1974,Ninio1975,McKeithan1995}, and it is a non-equilibrium mechanism in the strict sense: switch off the energy supply and the discrimination vanishes.

More generally, living systems do not sit at equilibrium. Prigogine and co-workers showed that ordered structures emerge spontaneously in systems held far from equilibrium by a continuous flux of energy and matter~\cite{Prigogine1977}, and England extended the argument to show that groups of molecules can organize into configurations that dissipate energy more effectively~\cite{England2015}. The attractive forces that drive assembly are sculpted by that flux, and the resulting non-equilibrium steady state is exquisitely sensitive to perturbation.

The argument runs from the equilibrium free energy to the equations of motion and the residence time they imply, then through two consequences of that time: kinetic selectivity under detailed balance, worked out for targeted nanoparticles under flow, and discrimination beyond the equilibrium ceiling once detailed balance is broken. The second of these is developed along the valency axis, because the amplification a driven cascade supplies turns out to depend qualitatively on what kind of lifetime it reads. At one end sits the T cell, a contact of unit valency whose single-bond lifetime varies with ligand chemistry but barely with ligand density; there the proofreading exponent multiplies a slope of order one. At the other sits the multivalent particle, whose whole-contact lifetime is already exponential in receptor density through the Kramers relation; there the same exponent multiplies the engagement number, and the two amplifications compound. We therefore establish the mechanism on the single bond, where it is experimentally settled, and then ask what changes when the contact is multivalent, where it is not.

\section{The Equilibrium Starting Point}\label{sec:background}

We summarize here, without derivation, the equilibrium results the dynamics will need. Full derivations are given elsewhere~\cite{PATarxiv}; the point of this section is to make the present paper self-contained and to fix notation.

\subsection{The medium licenses an overdamped description}

Every interaction between two objects in water is a solvent-averaged interaction, a potential of mean force rather than an energy~\cite{Kirkwood1935,Roux1995}. That description is dynamically meaningful only if the integrated-out coordinates equilibrate quickly compared with the motion of the retained ones, and at the nanoscale they do, by a wide margin. Water rearranges in $1$ to $5$~ps, while a $5$-nm object with diffusivity $D\simeq49\ \mu$m$^2$\,s$^{-1}$ in water of viscosity $0.89$~mPa\,s needs about $10$~ns to shuffle a single nanometer. The solvent is therefore always at equilibrium with respect to the instantaneous configuration of the units, and the potential of mean force acts as a genuine potential governing their motion. The same estimate gives a Reynolds number so small that inertia never matters: nothing at this scale coasts. Motion is overdamped, and that single fact is why the dynamics below are described by a Smoluchowski equation rather than by Newton's.

\subsection{The multivalent unit and its free energy}

A \emph{multivalent unit} is a scaffold carrying $Z$ populations of binding moieties, each of valency $N_{{{l}},\zeta}$ and per-bond free energy $\varepsilon_\zeta$, wrapped in a non-specific repulsive layer that we call generically the \emph{steric toll}. Two such units in contact are described by their surface separation $h$ and by the \emph{engagement vector} $\mathbf{k}=(k_1,\ldots,k_Z)$ counting the bonds formed by each moiety type. The number of microstates realizing a given $\mathbf{k}$ is the \emph{topology kernel} $\Omega(\mathbf{k})$, and its logarithm is the combinatorial entropy that Kitov and Bundle isolated as the statistical origin of the multivalency effect~\cite{Kitov2003}. For a single moiety type the kernel is exactly the number of $k$-matchings of a bipartite reachability graph between ligands and receptors, which for a flexible unit against $N_{\rm r}$ reachable receptors reduces to $\Omega(k)=\binom{N_{{l}}}{k}\binom{N_{\rm r}}{k}k!$ and, in the mean-field limit, to $\binom{N_{{l}}}{k}b^k$ with $b$ the per-ligand receptor availability.

Assuming bond free energies add, the free energy of a configuration is
\begin{equation}\label{eq:additive}
G^{(0)}(\mathbf{k},h)=\sum_{\zeta=1}^{Z}k_\zeta\,\varepsilon_\zeta(h)-\kt\ln\Omega(\mathbf{k})+W_{\rm rep}(h),
\end{equation}
where the first term on the right sums the free energies of the bonds formed, each engaged moiety of type $\zeta$ contributing $\varepsilon_\zeta(h)$; the second is the combinatorial entropy of the $\Omega(\mathbf{k})$ ways of forming them; and $W_{\rm rep}(h)$ is the \emph{steric toll}: a smooth, non-specific, entropically generated repulsion that holds the two surfaces apart and, by charging a price every partner must pay, sets the threshold for recognition. The toll's microscopic origin is system-specific while its role in the theory is invariant, and it is worth naming the instances, because the generalization matters for what follows. In nanomedicines it is typically the stabilization layer used to prevent protein opsonization, often made of a polymer brush of poly(ethylene glycol) or polysaccharide whose osmotic and elastic response is the classical grafted-brush repulsion~\cite{Alexander1977,Milner1988,Jokerst2011,Suk2016}; throughout we reserve the word corona for the adsorbed protein layer such a brush exists to prevent, and call the repulsive layer itself the steric toll. On a cell the tolls stack from the membrane outward. At the bilayer itself it is \emph{membrane crowding}, the two-dimensional pressure of a receptor sitting shoulder to shoulder with a dense field of other proteins and glycans~\cite{Guigas2016}. Above it is the \emph{glycocalyx}, the sheath of membrane glycans, proteoglycans and disordered protein tails that coats every eukaryotic surface~\cite{Weinbaum2007,Mockl2020}, thickening at an epithelium into the mucin mesh. And in the cytosol it is the osmotic pressure of three-dimensional \emph{macromolecular crowding}, at concentrations that routinely exceed $300$~mg\,mL$^{-1}$~\cite{Zhou2008}. To these the residual electrostatic double layer and the hydration force add their share. We carry them in a single $W_{\rm rep}$ because in the threshold they are interchangeable: none carries contrast in the sense of Eq.~\eqref{eq:discrim}, so none can recognize, and all must be paid. The steric toll is the gate; the specific bonds are the key.

A second class of non-specific steric physics depends on engagement rather than on separation alone, and so cannot be folded into $W_{\rm rep}(h)$. These are the many-body elements, and they enter through a coupling matrix $\mathbf{J}$. Scaffold connectivity removes configurational entropy, because the first bond localizes the whole unit and confiscates freedom the others were exploiting. The interposed layers compress as bonds pull the surfaces together, so the cost of the $(k{+}1)$-th bond grows with the number already formed. Bulk macromolecular crowding, whose pairwise depletion between two units is attractive, contributes an irreducible three-body term that is repulsive and switches on above a crowder-to-unit size ratio of about $0.155$~\cite{Asakura1954,PerezAngel2025}; membrane curvature couples three or more inclusions non-additively besides. Membrane crowding enters here too, and in a way worth stating explicitly, because it is specific to recognition at a surface. When a ligand docks on its receptor $X$ it is brought into contact with whatever sits beside $X$ on the target membrane, and its interaction with those bystander species $Y$ is non-specific and usually repulsive. This surface-crowding penalty renormalizes the per-bond free energy upward, and because each bond formed packs the interface more tightly and so raises the repulsion the next ligand must overcome, it contributes a positive, engagement-dependent coupling to $\mathbf{J}$. It is the lateral, membrane-bound counterpart of bulk crowding. Collecting these into a symmetric coupling matrix $\mathbf{J}$ and truncating at leading non-additive order,
\begin{equation}\label{eq:withMB}
\mathcal{G}(\mathbf{k},h)=G^{(0)}(\mathbf{k},h)+\sum_{\zeta\le\zeta'}J_{\zeta\zeta'}\,k_\zeta k_{\zeta'}+\mathcal{O}(k^3),
\end{equation}
with $J<0$ where cooperative mechanisms dominate (partner recruitment, lateral partner attraction~\cite{Curk2025}) and $J>0$ where anti-cooperative ones do (crowding depletion, surface-crowding repulsion from receptor neighbors, scaffold connectivity, compression of the interposed layers). The diagonal $J_{\zeta\zeta}>0$ is what makes engagement self-limiting: if the cost of the $(k{+}1)$-th bond rises with $k$, then beyond some engagement the next bond is not worth forming, and recognition acquires a ceiling as well as a floor. We shall see that ceiling appear explicitly in the engagement landscape of Sec.~\ref{sec:kinetic}.

Writing the combined bond and coupling energy of an engagement state as
\begin{equation}\label{eq:configen}
E_{\mathbf{k}}(h)\equiv\sum_\zeta k_\zeta\varepsilon_\zeta(h)+\sum_{\zeta\le\zeta'}J_{\zeta\zeta'}k_\zeta k_{\zeta'},
\end{equation}
summing Boltzmann weights over every engagement state gives the unit-to-unit potential of mean force,
\begin{equation}\label{eq:master}
W_{AB}(h)=W_{\rm rep}(h)-\kt\ln\sum_{\mathbf{k}}\Omega(\mathbf{k})\,e^{-\beta E_{\mathbf{k}}(h)},
\end{equation}
with $\beta\equiv1/\kt$ and $\Omega(\mathbf{k})$ the topology kernel. Equation~\eqref{eq:master} is the equilibrium theory in one line, and it is the potential in which the dynamics of Sec.~\ref{sec:smol} take place.

\subsection{Two results we shall lean on}

Two consequences of the equilibrium theory are needed below, and both are rigorous rather than modeled.

The first is the \emph{equilibrium discrimination ceiling}. Because binding constants are exponential in free energy, the ratio of bound populations for a cognate partner and a typical non-cognate one is
\begin{equation}\label{eq:discrim}
\frac{K_{\rm cognate}}{\langle K_{\rm non\text{-}cognate}\rangle}=e^{\beta\,\Delta\varepsilon},
\end{equation}
where $\Delta\varepsilon$ is the free-energy contrast. The scale is worth committing to memory: $\Delta\varepsilon=2.3\,\kt$ buys one decade of discrimination, $4.6\,\kt$ buys two, $9.2\,\kt$ buys four. No equilibrium mechanism, however elaborate, can do better than Eq.~\eqref{eq:discrim} with the full free-energy difference inserted. Multivalency raises the effective $\Delta\varepsilon$ by summing contrasts over bonds; it does not escape the ceiling.

The second is \emph{unimodality}. For a single moiety type the contact partition function $\Xi(x)=\sum_k\Omega(k)x^k$, $x=e^{-\beta\varepsilon}$, is the matching generating polynomial of the reachability graph, and Heilmann and Lieb proved that such a polynomial has only real, non-positive roots~\cite{HeilmannLieb1972}. Real-rootedness implies that the kernel is log-concave,
\begin{equation}\label{eq:logconcave}
\Omega(k)^{2}\ \ge\ \Omega(k-1)\,\Omega(k+1),
\end{equation}
so the equilibrium distribution of bond number is unimodal for every architecture: there is always a single most probable engagement $k^\star$, never two competing ones. Put plainly, the count of ways to form $k$ bonds has one peak as $k$ grows, so a contact settles on a single preferred number of bonds instead of flickering between a weakly and a strongly engaged state. This is the formal reason the landscapes below have one bound minimum rather than several, and it is what justifies characterizing a contact by $k^\star$ and by the barrier separating $k^\star$ from $k=0$.

That is all the equilibrium input we need. It is worth pausing on how little it assumes: two objects, some binding groups, a repulsive layer, and water. What it cannot do is tell us how long anything takes.

\section{From Master Equation to Smoluchowski}\label{sec:smol}

There are two natural coordinates, and they call for two different but equivalent descriptions. Engagement $\mathbf{k}$ is a discrete integer count, so its evolution is naturally a jump process. Separation $h$ is continuous, so its evolution is naturally a diffusion in a potential.

Let $P(\mathbf{k},t)$ be the probability of engagement $\mathbf{k}$ at time $t$. Conservation of probability over the discrete state space gives the master equation
\begin{equation}\label{eq:mastereq}
\frac{\partial P(\mathbf{k},t)}{\partial t}
=\sum_{\mathbf{k}'}\bigl[\,W(\mathbf{k}'\!\to\!\mathbf{k})P(\mathbf{k}',t)-W(\mathbf{k}\!\to\!\mathbf{k}')P(\mathbf{k},t)\bigr],
\end{equation}
and the rates are not free to choose. Any closed system relaxing to the equilibrium of Eq.~\eqref{eq:withMB} must satisfy detailed balance with respect to it,
\begin{equation}\label{eq:detbal}
\frac{W(\mathbf{k}\!\to\!\mathbf{k}')}{W(\mathbf{k}'\!\to\!\mathbf{k})}
=\exp\!\bigl[-\beta\bigl(\mathcal{G}(\mathbf{k}')-\mathcal{G}(\mathbf{k})\bigr)\bigr].
\end{equation}
For single-bond formation and rupture a convenient parametrization consistent with Eq.~\eqref{eq:detbal} is
\begin{equation}\label{eq:rates}
\begin{aligned}
W(k\!\to\!k{+}1)&=k_{\rm on}\,(N_{\rm r}-k)\,\frac{\Omega(k{+}1)}{\Omega(k)}\,e^{-\beta\Delta\mathcal{G}^{\ddagger}_{\rm on}},\\[4pt]
W(k\!\to\!k{-}1)&=k_{\rm off}\,k ,
\end{aligned}
\end{equation}
in which the topology kernel appears explicitly as a ratio, so that architecture enters the kinetics as well as the thermodynamics. Equation~\eqref{eq:detbal} is a constraint rather than a modeling choice, and rates that violate it describe a system driven by an external energy source. That is precisely what kinetic proofreading requires, and we return to it in Sec.~\ref{sec:proofreading}.

For the continuous coordinate the overdamped limit of the Fokker--Planck equation is the Smoluchowski equation~\cite{Smoluchowski1917,Risken1989},
\begin{equation}\label{eq:smol}
\frac{\partial P(h,t)}{\partial t}
=\frac{\partial}{\partial h}\!\left[D(h)\!\left(\frac{\partial P}{\partial h}+\beta P\,\frac{\partial W_{AB}}{\partial h}\right)\right]
\equiv-\frac{\partial \mathcal{J}}{\partial h},
\end{equation}
with $W_{AB}$ the potential of mean force of Eq.~\eqref{eq:master}, $D(h)$ the position-dependent relative diffusivity and $\mathcal{J}$ the probability flux. Equation~\eqref{eq:smol} may be rewritten in a manifestly detailed-balanced form,
\begin{equation}\label{eq:smol2}
\frac{\partial P}{\partial t}=\frac{\partial}{\partial h}\left[D\,e^{-\beta W_{AB}}\frac{\partial}{\partial h}\Bigl(e^{\beta W_{AB}}P\Bigr)\right],
\end{equation}
from which the zero-flux stationary solution is read off immediately as $P_{\rm eq}(h)\propto e^{-\beta W_{AB}(h)}$. Equilibrium statistical mechanics is therefore recovered as the long-time limit, and everything in Sec.~\ref{sec:background} is contained in the dynamics rather than replaced by it. A stationary state with $\mathcal{J}\neq0$, by contrast, is a non-equilibrium steady state sustained only by driving.

The quantity we actually want is a timescale. Integrating Eq.~\eqref{eq:smol} twice with an absorbing boundary gives the mean first-passage time from $h_0$ to a target at $h_t$,
\begin{equation}\label{eq:mfpt}
\tau(h_0\!\to\!h_t)=\int_{h_t}^{h_0}\!\frac{\mathrm{d}y}{D(y)}\,e^{\beta W_{AB}(y)}\!\int_{y}^{\infty}\!\mathrm{d}x\;e^{-\beta W_{AB}(x)},
\end{equation}
and when the landscape has a well of depth $\Delta\mathcal{G}^{\ddagger}$ separated from the outside by a sharp barrier, the double integral is dominated by its endpoints and reduces to the Kramers form~\cite{Kramers1940,Hanggi1990}
\begin{equation}\label{eq:kramers}
\tau\simeq\tau_0\,e^{\beta\Delta\mathcal{G}^{\ddagger}},
\qquad \tau_0\sim1\ \text{ns for nanoscale units.}
\end{equation}

Equation~\eqref{eq:kramers} is the single most consequential result in this paper, because it says that times depend \emph{exponentially} on free energies. A free energy that changes by a factor of two changes a rate by its exponential, and since the free energies in question are tens of $\kt$, the rates span many orders of magnitude. Recognition thresholds that look modest on a thermodynamic plot become ferociously sharp when read as lifetimes.

\section{Kinetic Selectivity: Reading Density Through Lifetime}\label{sec:kinetic}

\begin{figure*}[t]
\centering
\includegraphics[width=\textwidth]{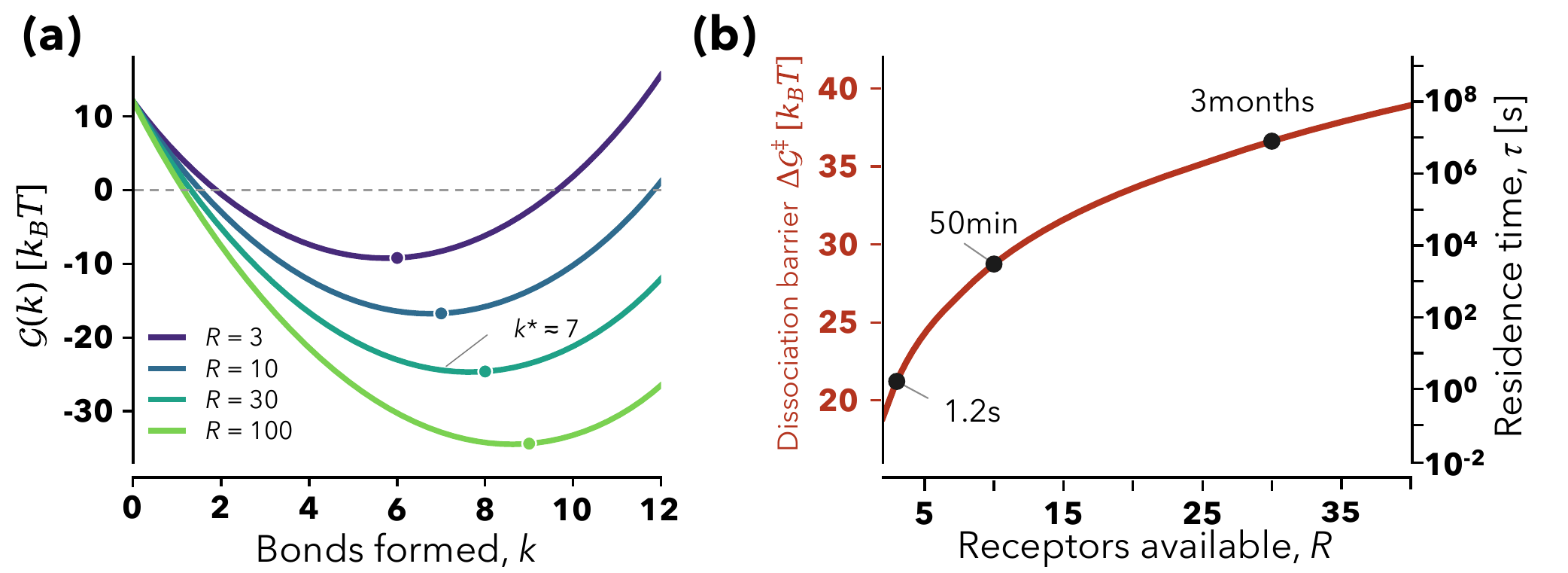}
\caption{Dynamics of a multivalent contact, computed from Eq.~\eqref{eq:withMB} with $N_{{l}}=12$, $\varepsilon=-4\,\kt$, steric toll $W_{\rm rep}=12\,\kt$ and self-coupling $J=0.45\,\kt$. (a)~The engagement landscape $\mathcal{G}(k)$ for receptor numbers $N_{\rm r}=3,10,30,100$. Approach must first pay the steric toll at $k=0$, which acts as the transition state; the bound minimum then deepens with $N_{\rm r}$. Because $J>0$, engagement self-limits near $k^\star\approx7$ however many receptors are offered. (b)~Dissociation barrier $\Delta\mathcal{G}^{\ddagger}$ (left) and the resulting residence time $\tau$ from Eq.~\eqref{eq:kramers} (right) against receptor number. A tenfold change in $N_{\rm r}$ moves the lifetime of the contact from seconds to months.}
\label{fig:dynamics}
\end{figure*}

Fig.~\ref{fig:dynamics}(a) shows the engagement landscape $\mathcal{G}(k)$ of Eq.~\eqref{eq:withMB} for a representative construct, a twelve-valent unit with per-bond free energy $\varepsilon=-4\,\kt$, steric toll $W_{\rm rep}=12\,\kt$ and anti-cooperative self-coupling $J=0.45\,\kt$, and its shape is worth reading carefully because every feature was predicted rather than assumed. At $k=0$ the free energy sits at the full steric toll, because the units are in contact but no bond has yet formed: the repulsive layer that equilibrium theory shows to be obligatory now doubles as the \emph{transition state} of the association reaction. Bonds then form, each paying for itself and for the combinatorial entropy it unlocks, until the anti-cooperative coupling $J>0$ makes further bonds unprofitable. The minimum sits near $k^\star\approx7$ and, strikingly, barely moves as the receptor supply is increased tenfold. This is the ceiling on engagement that steric compression and scaffold connectivity predict, and it appears here without any additional assumption. That the minimum is unique rather than one of several is guaranteed by the log-concavity of Eq.~\eqref{eq:logconcave}.

Fig.~\ref{fig:dynamics}(b) converts landscape depth into time using Eq.~\eqref{eq:kramers}, and the numbers are startling. With three receptors available the contact lives about a second. With ten it lives close to an hour. With thirty it lives months. A factor of ten in receptor number spans nearly seven orders of magnitude in residence time. Nothing in the thermodynamics changes character across that range; the free energy simply grows smoothly. It is the exponential in Eq.~\eqref{eq:kramers} that turns a smooth thermodynamic gradient into a switch.

This has a direct practical consequence. If a construct encounters a surface only briefly, as under flow, then the relevant question is whether it can form enough bonds during the encounter to survive it. Since the number of bonds achievable in a fixed time depends on receptor density through the rates of Eq.~\eqref{eq:rates}, and the survival probability depends exponentially on the number of bonds, selectivity measured kinetically can be considerably sharper than selectivity measured at equilibrium. Experiments on multivalent polymers have found exactly this, with initial association more strongly density-selective than the final equilibrium state~\cite{Ravnik2026}. Design for a brief encounter and design for a long incubation are therefore different problems with different optima, a point we make concrete in the next section.

It is worth separating the two things Fig.~\ref{fig:dynamics} shows, because they are often conflated. The \emph{position} of the threshold, the receptor number at which the bound minimum dips below zero, is a thermodynamic quantity set by the balance between the toll and the combinatorial payment. The \emph{sharpness} of the threshold read as a lifetime is a kinetic quantity, set by the exponential of Eq.~\eqref{eq:kramers}. Equilibrium theory places the threshold; time decides how decisive it is.

\section{Multivalent Nanoparticles: Reading Receptor Density Under Flow}\label{sec:nano}

A multivalent unit, in synthetic form, is any scaffold that carries many copies of a ligand: a polymersome, a micelle, a liposome, a dendrimer, a solid nanoparticle. Whatever the chemistry, it maps term for term onto the unit of Sec.~\ref{sec:background}, and the equilibrium part of that mapping has already been validated experimentally~\cite{Tian2020,Liu2020}. The scaffold is the body, the surface ligands are the moieties of valency $N_{{l}}$ and per-bond free energy $\varepsilon$, and the poly(ethylene glycol) brush is the steric toll $W_{\rm rep}$ in its synthetic form. That brush is not a finishing touch. It is needed in the first place to hold off the protein corona and the non-specific interactions that would otherwise decide where the particle ends up, and it does a second job that matters just as much: by setting how far each ligand must reach to find a receptor, the brush tunes the single-bond free energy down to the weak value superselectivity requires, as we showed in designing precision nanomedicines~\cite{Tian2020}. A strong bond sticks to everything. Superselectivity is built from many deliberately weak ones, and the brush is how that weakness is dialed in. The glycocalyx on the target surface adds its own toll, so $W_{\rm rep}$ is the sum of both layers the particle must compress before any bond forms. The design parameters, size, brush length and grafting density, tether length, ligand affinity and ligand number, are the descriptors of the functional~\cite{Tian2020}, and tuning them sets the receptor density at which the particle switches from loose to bound.

At equilibrium this produces superselectivity, and the theory reproduces the measured behavior. Using the single-type reduction of Eq.~\eqref{eq:master}, $\Xi=(1+b)^{N_{{l}}}$ with $b$ the per-ligand receptor availability, the selectivity exponent $\alpha\equiv \mathrm{d}\ln\theta/\mathrm{d}\ln\rho$ of the bound fraction $\theta$ climbs from the monovalent ceiling $\alpha\le1$ as valency rises. For a steric toll $W_{\rm rep}=9\,\kt$ and the per-ligand free energy tuned in each case to place the threshold at a common receptor density, we compute $\alpha_{\max}=1.0$ for a monovalent ligand, $6.9$ at $N_{{l}}=10$, $10.0$ at $N_{{l}}=30$ and a plateau near $11$ by $N_{{l}}=50$. The plateau is not a numerical artifact: it is the engagement ceiling of the anti-cooperative coupling, the same $J>0$ that pins $k^\star$ in Fig.~\ref{fig:dynamics}(a). A particle cannot be made arbitrarily selective by adding ligands, because the steric layer it must compress to engage them pushes back. Read as discrimination against a threefold difference in receptor density, the monovalent ligand is held at the ceiling of Eq.~\eqref{eq:discrim}, exactly the $3.0$-fold density ratio and no better, while a multivalent particle of $N_{{l}}=10$ in the superselective window reads the same threefold difference as roughly $25$-fold. This is the quantitative content of designing a carrier to read receptor density rather than receptor identity~\cite{Tian2020,Liu2020,Dubacheva2023}.

What the equilibrium picture misses is that in an organism a particle never simply incubates against its target. It is carried there, and transport in tissue is a competition between diffusion and advection, the bulk flow of blood and interstitial fluid. How long a particle dwells against a given surface therefore depends on where that surface sits. In a poorly perfused tumour, where the vessels are leaky and sluggish and the raised interstitial pressure stalls convection, a particle can linger for minutes to hours; we have used exactly that long residence to carry paclitaxel-loaded polymersomes into peritoneal carcinomatosis~\cite{SimonGracia2016}. At the brain endothelium the opposite holds: a particle is swept past the luminal surface in well under a second, and only a contact that survives the transit is read~\cite{Joseph2020}. The dwell time is not a property of the particle. It is set by the target.

This is the kinetic-selectivity argument of Sec.~\ref{sec:kinetic} in its most literal form. Taking the single-type construct at $N_{{l}}=30$ and converting the bound-minimum depth into a residence time through Eq.~\eqref{eq:kramers}, the contact lifetime moves from well under a millisecond at the reference density, to about a minute at threefold that density, to effectively permanent at tenfold. Against a brain-endothelial dwell of $0.1$ to $1$~s the receptor-density axis is cut sharply: below the cutoff the contact is shorter than the transit and the particle is washed off, above it the contact outlasts the transit and the particle is kept. Against the minutes-to-hours dwell of a sluggish tumour the same particle binds far down its density curve, close to the equilibrium limit. The capture threshold sits where the lifetime curve crosses the dwell time, and because that curve is exponential in free energy, the threshold is far sharper than the equilibrium bound-fraction curve through the same point. This resolves something we had seen but not fully explained: constructs tuned for superselective binding in vitro do better in vivo than the static assay predicts~\cite{Tian2020,Joseph2020}, because the living target adds a kinetic filter the plate does not have.

\begin{figure*}[t]
\centering
\includegraphics[width=\textwidth]{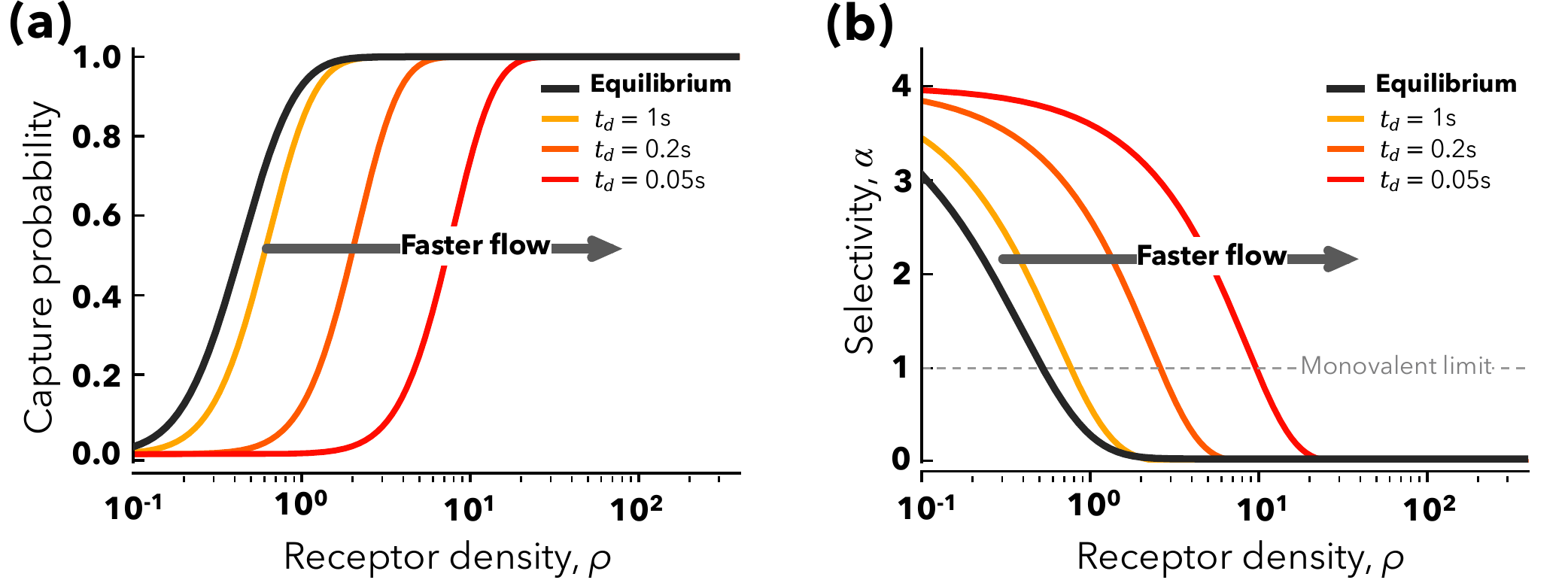}
\caption{Kinetic superselectivity of the Fig.~\ref{fig:dynamics} particle under flow, from the master equation~\eqref{eq:mastereq} with rates~\eqref{eq:rates}, $N_{{l}}=12$, and capture defined as reaching $k_c=4$ bonds within the dwell time $t_d$ starting from the unbound state. (a)~Capture probability against receptor density for a long incubation (equilibrium) and for three dwell times. As the flow speeds up and $t_d$ shortens, the density threshold for capture shifts to higher receptor density and the curve steepens. (b)~The corresponding selectivity exponent $\alpha=\mathrm{d}\ln\theta/\mathrm{d}\ln\rho$. The flow-limited curves sit above the equilibrium one and extend the superselective window ($\alpha>1$) to higher density, with $\alpha_{\max}$ rising from $3.1$ at equilibrium toward the ceiling $k_c$ as the encounter shortens.}
\label{fig:flow}
\end{figure*}

This is not only a qualitative argument; the same master equation that produced Fig.~\ref{fig:dynamics} delivers it quantitatively. We take the $N_{{l}}=12$ particle of Fig.~\ref{fig:dynamics}, start it unbound at $k=0$, evolve the engagement under the detailed-balanced rates of Eq.~\eqref{eq:rates}, and identify \emph{capture} with reaching a critical $k_c=4$ bonds, the number needed to resist the hydrodynamic drag, within the flow dwell time $t_d$. Fig.~\ref{fig:flow} shows the result, and as the flow accelerates and $t_d$ falls from $1$~s to $50$~ms two things happen together. The receptor density at which the particle is captured climbs more than an order of magnitude (Fig.~\ref{fig:flow}a), because a shorter encounter demands a faster accumulation of bonds and hence more receptors; and the selectivity exponent rises from its equilibrium value $\alpha_{\max}\simeq3.1$ to $\simeq4.0$ (Fig.~\ref{fig:flow}b), approaching the ceiling $k_c$ set by the number of bonds the capture requires. The transient, flow-limited state is therefore \emph{more} density-selective than the equilibrium state the particle would reach given unlimited time, which is exactly the kinetic superselectivity reported for multivalent polymers~\cite{Ravnik2026}. The reading is pointed: the same particle, unchanged in chemistry, reads a different and sharper receptor-density threshold depending only on the shear it is delivered under.

Two design consequences follow, and both are testable with the kind of flow and static binding assays already used for these systems. First, the optimal ligand density under flow is not the optimal ligand density at equilibrium. A static incubation rewards whatever maximizes equilibrium avidity; a brief encounter rewards fast engagement, which weights on-rate and the number of bonds reachable in the dwell time, Eq.~\eqref{eq:rates}, over the ultimate depth of the well. A carrier tuned on a binding plate and then deployed in vivo is optimized for the wrong problem. Second, as Fig.~\ref{fig:flow} makes concrete, the same particle reads a different and sharper receptor-density cutoff purely by the flow regime it is delivered in, without any change to its chemistry, the kinetic counterpart of the superselectivity sweet spots mapped at equilibrium~\cite{Tian2020}. Everything to this point respects detailed balance. The particle thread resumes in Sec.~\ref{sec:uptake}, once the driven machinery has been set up on the simpler single-bond case.

\section{Breaking Detailed Balance: Proofreading a Single Bond}\label{sec:proofreading}\label{sec:tcell}

Everything so far respects detailed balance, Eq.~\eqref{eq:detbal}, and is therefore held under a hard ceiling: at equilibrium two competing targets can be told apart by no more than the ratio of their Boltzmann factors, Eq.~\eqref{eq:discrim}. A difference of a few $\kt$ buys one or two decades of discrimination and not a decade more. Several of the discriminations cells actually perform run past that ceiling, and they manage it by spending energy. We develop the mechanism here on the simplest case the framework contains, a contact of unit valency, and then ask in Sec.~\ref{sec:uptake} what changes when the lifetime being read belongs to a whole multivalent contact rather than to one bond. The two are the ends of the valency axis on which this paper is built, and the amplification differs qualitatively between them.

The immune system spends that energy on a single bond, and spends more of it than any synthetic particle does, which makes it the natural place to establish the mechanism. In the language of Sec.~\ref{sec:background} a T cell is a multivalent unit at the $N_{{l}}\to1$ boundary, one to a few T cell receptors engaging peptide-loaded MHC within the contact, with per-bond free energies in the range $-6$ to $-10\,\kt$. It is the limiting case that calibrates the framework: whatever amplification the driven machinery supplies here, it supplies without any help from multivalency. A T cell must pick a peptide from a pathogen out of the tens of thousands of self peptides displayed on the same major histocompatibility complex molecules, and a single amino-acid substitution can turn a fully activating peptide inert~\cite{Evavold1991,Davis1998}. The binding free energies of agonist and non-agonist peptides differ by only a few $\kt$, so Eq.~\eqref{eq:discrim} allows equilibrium at most one or two decades of discrimination, far below what is observed. The one property that sets such a contact apart from an antibody or a nanoparticle is that its selectivity is \emph{kinetic}, proofread and driven, rather than thermodynamic.

The cell is not asking how tightly the complex binds. It is asking whether the complex lasts long enough for the machinery to commit, and that question has a classical answer. Hopfield and Ninio showed, and McKeithan brought to immune recognition~\cite{Hopfield1974,Ninio1975,McKeithan1995,Francois2013}, that a chain of irreversible, energy-consuming steps placed between binding and commitment turns a small difference in contact lifetime into a large difference in outcome. Insert $N$ such steps, each proceeding at rate $k_p$, and let the contact be lost at rate $1/\tau$ at every stage. The chance of surviving one step is the race between progression and loss, $k_p/(k_p+1/\tau)$, and because the steps are independent the chance of clearing all $N$ is
\begin{equation}\label{eq:proof}
P(\tau)=\left(\frac{k_p\tau}{1+k_p\tau}\right)^{\!N}.
\end{equation}
Dividing the survival of two contacts with lifetimes $\tau_1>\tau_2$ gives the discrimination,
\begin{equation}\label{eq:proofdisc}
\frac{P(\tau_1)}{P(\tau_2)}=\left(\frac{\tau_1}{\tau_2}\cdot\frac{1+k_p\tau_2}{1+k_p\tau_1}\right)^{\!N}\;\xrightarrow[k_p\tau\ll1]{}\;\left(\frac{\tau_1}{\tau_2}\right)^{\!N},
\end{equation}
a power law in the lifetime ratio whose exponent is the number of driven steps. A tenfold difference in lifetime is worth only about $2.3\,\kt$, so at most tenfold discrimination at equilibrium, but it becomes $10^N$ once the cascade runs: three steps give a thousandfold, five a hundred thousandfold. The free-energy landscape of Sec.~\ref{sec:background} has not gone anywhere; it still sets the lifetime $\tau$ through the Kramers relation~\eqref{eq:kramers}. What the cascade adds is a readout exquisitely sensitive to that lifetime, paid for in metabolic energy. The thermodynamics sets the raw signal, the driven machinery sets the gain. The price is sensitivity, since $P\le1$ falls with $N$: a cell that proofreads more stringently needs more of its target to respond at all.

\subsection{Specificity is manufactured, not measured}

\begin{figure*}[t]
\centering
\includegraphics[width=\textwidth]{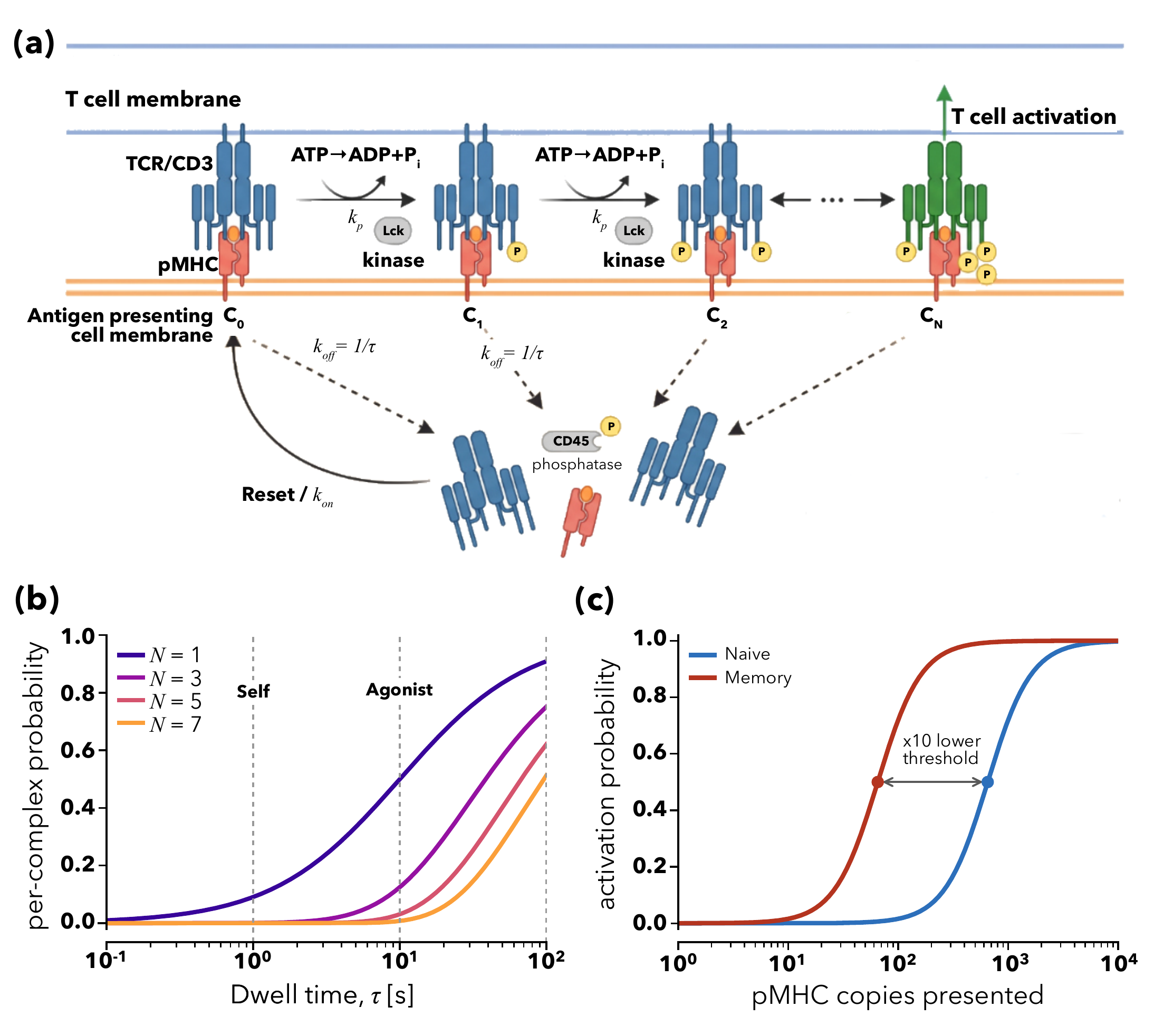}
\caption{T cell recognition as a driven, non-equilibrium process. (a)~Schematic of kinetic proofreading: a bound TCR--pMHC complex passes through $N$ energy-consuming modification steps (rate $k_p$, each irreversible and ATP-costing) before it is licensed to signal, and if the complex unbinds (rate $k_{\rm off}=1/\tau$) the cascade resets to the start. (b)~Kinetic proofreading, Eq.~\eqref{eq:proof}, with $k_p=0.1$~s$^{-1}$. Increasing the number of steps sharpens discrimination between a short-lived self complex and a long-lived agonist complex, at the cost of overall signal amplitude. (c)~Naive against memory T cells. Lowering the downstream activation threshold shifts the antigen-density requirement by an order of magnitude while leaving the proofreading machinery, and therefore the specificity, untouched.}
\label{fig:tcell}
\end{figure*}

Eq.~\eqref{eq:proof} applied to a single bond is quantitatively adequate for T cells, as experiment and modeling have shown~\cite{AltanBonnet2005,Francois2013}. The T cell does not measure how tightly the complex binds; it measures how long the complex survives~\cite{Aleksic2010,Govern2010}, by requiring it to pass through a series of energy-consuming modification steps before it is allowed to signal. Fig.~\ref{fig:tcell}(b) shows the effect, computed from Eq.~\eqref{eq:proof} with $k_p=0.1$~s$^{-1}$ for a short-lived self complex of dwell time $1$~s against a long-lived agonist complex of $10$~s. With a single step, a tenfold difference in dwell time yields sixfold discrimination, roughly the equilibrium result. With three steps it yields $166$-fold, with five $5\times10^3$-fold and with seven $1.5\times10^5$-fold. Specificity is manufactured, not measured, and it is paid for in ATP. That this suffices is remarkable given the numbers: a handful of agonist pMHC among $10^5$ self ligands is enough to trigger a response~\cite{Irvine2002,Purbhoo2004}, within a contact whose architecture is itself organized by adhesion receptors into an immunological synapse~\cite{Grakoui1999,Dustin2014}.

Notice the trade-off visible in Fig.~\ref{fig:tcell}(b), because it is not optional. As $N$ increases the curves shift down as well as steepening: the probability that any complex completes the cascade falls. Specificity is bought with sensitivity. A cell that proofreads more stringently needs more antigen to respond at all. This tension is a genuine design constraint, and it sets up the final observation of this section.

\subsection{What memory changes, and what it does not}

Immune memory is usually described as a faster and stronger response to a previously encountered antigen. The framework lets us ask a sharper question: which parameter changes? There are two possibilities. Memory could improve the \emph{specificity}, by adding proofreading steps or lengthening agonist dwell times, or it could lower the \emph{threshold}, by making the downstream machinery easier to trigger. These are different interventions with different signatures.

Fig.~\ref{fig:tcell}(c) shows the second. Lowering the activation threshold, which corresponds biologically to pre-assembled signaling machinery, elevated adhesion receptor density and a relaxed costimulation requirement~\cite{Sallusto1999,Farber2014}, shifts the antigen-density response curve by roughly an order of magnitude: a memory cell responds to about $65$ peptide copies where a naive cell needs some $650$. The steepness of the curve, and therefore the discrimination between agonist and self, is unchanged.

This is the resolution of the sensitivity--specificity tension noted above, and it is rather elegant. A naive T cell must be stringent, because it has never seen the antigen and a false positive risks autoimmunity. It therefore sets a high threshold and accepts poor sensitivity. Once an antigen has been validated by a successful primary response, stringency can safely be relaxed for that clone specifically, and the threshold can drop. Memory, in this reading, is not better recognition. It is the same recognition machinery operating with its gate held lower, and the safety of doing so was established during the primary response. The specificity was never the thing that improved.

The two interventions are experimentally distinguishable, which is what makes the distinction more than a matter of interpretation. A change in specificity moves the \emph{slope} of the response against dwell time, and would show up as a change in the exponent $N$ of Eq.~\eqref{eq:proofdisc} or in the agonist lifetime $\tau_1$. A change in threshold moves the response curve \emph{sideways} along the antigen-density axis at fixed slope. Dose--response data on naive and memory clones against a graded series of altered peptide ligands would separate the two directly.
\section{Proofreading a Multivalent Contact: Endocytosis}\label{sec:uptake}

\begin{figure*}[t]
\centering
\includegraphics[width=\textwidth]{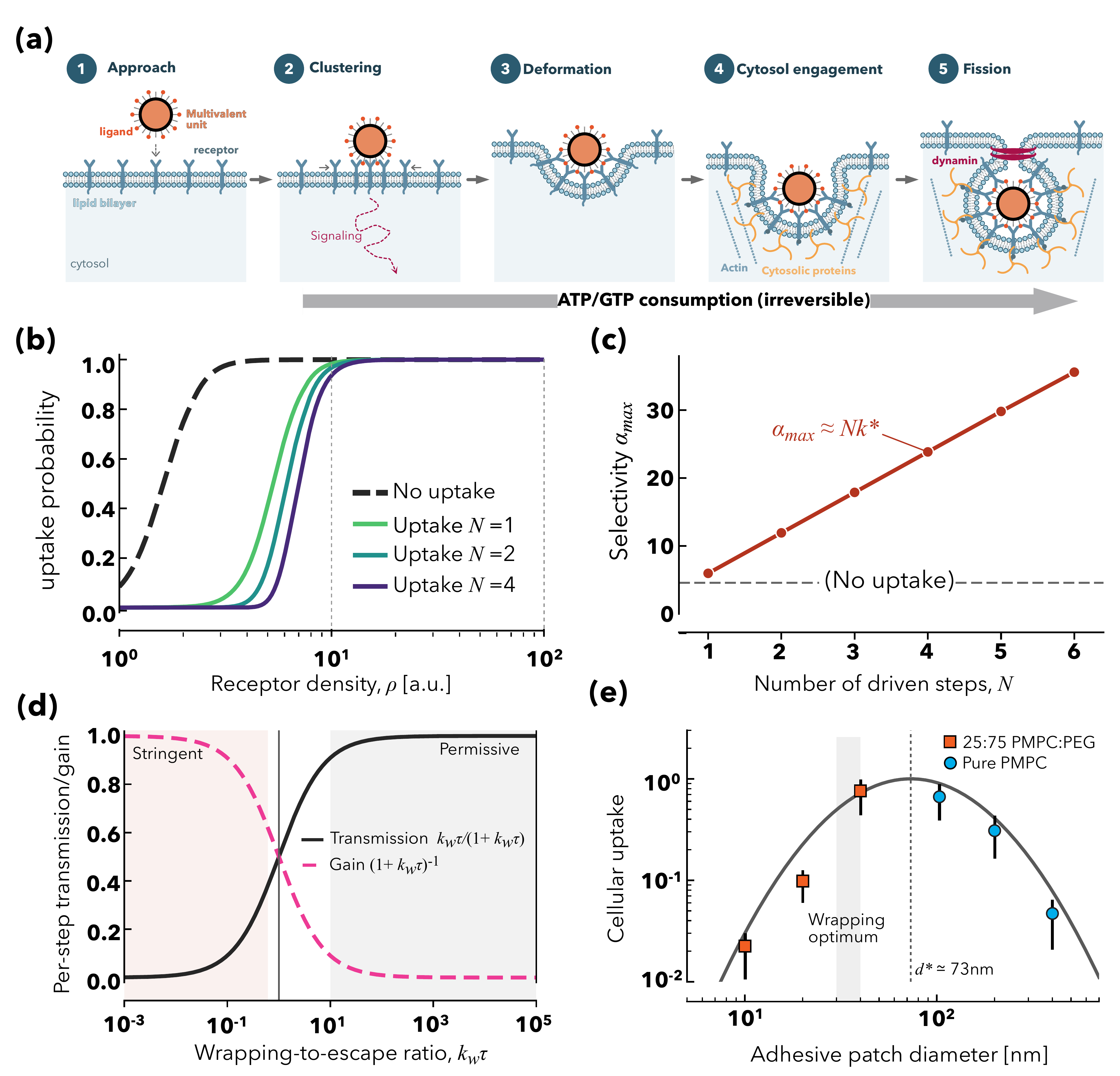}
\caption{Endocytosis as proofreading of the contact lifetime, for the $N_{{l}}=12$ particle of Fig.~\ref{fig:dynamics}. (a)~Schematic: a multivalent particle is engulfed through a sequence of largely irreversible, energy-consuming steps (receptor clustering, coat assembly, membrane wrapping, dynamin scission), and if it detaches (rate $1/\tau$) before scission the attempt resets. The contact lifetime $\tau(\rho)$ that the cascade reads is the one from Fig.~\ref{fig:dynamics}(b). (b)~Internalization probability, Eq.~\eqref{eq:uptake}, against receptor density for the equilibrium readout with the drive switched off and for the driven cascade with $N=1,2,4$ steps. Driven uptake reads a higher and sharper density threshold than equilibrium binding, because the contact must outlive the wrapping clock. (c)~The selectivity exponent grows linearly with the number of driven steps, $\alpha_{\max}\simeq N\,k^\star$, far above the equilibrium superselective value (dashed), the double amplification of the Kramers exponential by the proofreading power law. (d)~The regime is set by the single group $k_w\tau$, the wrapping rate against the escape rate. Each driven step transmits $k_w\tau/(1+k_w\tau)$ of the contacts reaching it (solid) and contributes $(1+k_w\tau)^{-1}$ of the equilibrium density sensitivity (dashed), so discrimination and throughput trade off across the crossover at $k_w\tau\simeq1$. Geometry enters through $k_w$ alone, so a construct can be moved across the crossover at fixed chemistry. (e)~The wrapping clock reads the adhesive patch rather than the vesicle, from the $24$~h uptake data of Ref.~\cite{Massignani2009} normalized by the number of vesicles offered. Pure PMPC vesicles (patch $=$ vesicle, $100$--$400$~nm) and $25{:}75$ PMPC:PEG vesicles (patch $=$ PMPC domain, $10$--$40$~nm) depend on vesicle diameter with opposite sign, slopes $-1.6$ and $+2.5$, yet collapse onto a single peaked curve when plotted against patch diameter (log-normal fit, $R^2=0.95$, maximum at $d^\star\simeq73$~nm), close to the independently estimated wrapping optimum (shaded).}
\label{fig:uptake}
\end{figure*}

Return now to the particle of Sec.~\ref{sec:nano}, and ask what the cascade of Sec.~\ref{sec:proofreading} does when the lifetime it reads is not that of a single bond but of a whole multivalent contact. Between the first bonds and the committed, internalized vesicle the cell runs through a sequence of largely one-way, energy-consuming events: receptors cluster, a clathrin or caveolin coat nucleates and grows, the membrane bends and wraps, and dynamin hydrolyzes GTP to pinch the vesicle off~\cite{DohertyMcMahon2009,Gao2005}. Scission is irreversible, and the whole cascade is driven, so deplete ATP and endocytosis halts. If the particle lets go before the membrane has wrapped far enough, the attempt unwinds and resets to the start. The cell is not asking how tightly the particle binds. It is asking whether the contact lasts long enough for the machinery to commit, which is the question Eq.~\eqref{eq:proof} answers, with the step rate now the rate $k_w$ at which the wrapping advances and the lifetime the whole-contact residence time $\tau(\rho)$ already computed in Fig.~\ref{fig:dynamics}(b),

\begin{equation}\label{eq:uptake}
P_{\rm uptake}(\rho)=\left(\frac{k_w\,\tau(\rho)}{1+k_w\,\tau(\rho)}\right)^{\!N}.
\end{equation}
Something now happens that a single-bond readout cannot produce, and it is the point of the section. The residence time is already exponential in the binding free energy, $\tau\sim e^{\beta\Delta\mathcal{G}^{\ddagger}}$ with $\Delta\mathcal{G}^{\ddagger}$ growing roughly as $k^\star\ln\rho$, so the proofreading exponent $N$ multiplies a slope that is already steep. Fig.~\ref{fig:uptake} makes it quantitative for the $N_{{l}}=12$ particle of Fig.~\ref{fig:dynamics}. The selectivity exponent of uptake grows linearly with the number of driven steps, $\alpha_{\max}\simeq N\,k^\star$, reaching $\simeq12$ at two steps and $\simeq24$ at four, while the equilibrium binding it would show with the drive switched off is capped near the superselective value $\alpha\simeq4.6$. Internalization also sets in at a \emph{higher} receptor density than binding alone (Fig.~\ref{fig:uptake}b): a particle can be comfortably bound at equilibrium and still never be taken up, because its contact does not outlive the wrapping clock.

What Eq.~\eqref{eq:uptake} says, stripped to essentials, is that a single dimensionless group decides everything: the ratio $k_w\tau$ of the wrapping rate to the escape rate. Each driven step transmits a fraction $k_w\tau/(1+k_w\tau)$ of the contacts that reach it, and adds a fraction $(1+k_w\tau)^{-1}$ of the equilibrium density sensitivity to the readout. The two are in direct opposition and the crossover sits at $k_w\tau\simeq1$ [Fig.~\ref{fig:uptake}(d)]. When $k_w\tau\gg1$ the machinery commits faster than the contact can break, every bound particle is internalized, and the cascade discriminates nothing: uptake inherits the equilibrium selectivity of binding and no more. When $k_w\tau\lesssim1$ most attempts unwind before scission, uptake falls, and each step multiplies the density exponent. Proofreading is therefore not something a cell either does or does not do. It is a regime, entered only when internalization is slow compared with escape.

This carries a consequence for design, because both rates depend in part on the particle. The escape rate is fixed by the contact free energy through Eq.~\eqref{eq:kramers}, but the wrapping rate is set by how much membrane must be bent and how far receptors must diffuse to cover the object~\cite{Gao2005,Chithrani2006}, and those are properties of shape and size rather than of chemistry. A construct can therefore be moved across the crossover at fixed ligand, fixed receptor and fixed cell, by geometry alone. Elongated or high-curvature objects, which need more membrane and more receptors per unit volume enclosed, should wrap slowly and sit on the stringent side; compact objects near the wrapping optimum should commit quickly and sit on the permissive side. The prediction that follows inverts the usual design rule: the geometry taken up least is the geometry that reads receptor density most sharply, while the compact particle chosen throughout the field precisely because it is internalized best is kinetically blind, its uptake a faithful copy of its binding. Running one geometric series across cells of graded receptor expression tests this, and needs no inhibitor, no transfection and no change of chemistry.

The same argument fixes what the cell contributes. Because $k_w$ belongs to the machinery and $\tau$ to the contact, a fast-internalizing phenotype is permissive and reads density poorly, while a slow one is stringent and reads it sharply but needs longer-lived contacts to take up anything at all. Phenotype selectivity in delivery is therefore not only a matter of which receptors a cell displays, but of how fast it eats, and the two can be played against each other.

Size supplies a second geometric handle on $k_w$, and an earlier study of the same copolymer shows how far it reaches. Internalization of PMPC-PDPA polymersomes by human dermal fibroblasts falls by up to two orders of magnitude as the vesicle diameter is raised from $100$ to $400$~nm, while PEG-PDPA vesicles of matched size are taken up an order of magnitude more slowly throughout~\cite{Massignani2009}. That work also supplies the drive-off control the mechanism requires: uptake is abolished at $4\,^\circ$C, almost entirely suppressed by chloroquine, and arrested close to the plasma membrane by nocodazole, so internalization here is unambiguously driven rather than spontaneous, and the particle that is merely bound is the particle that stays outside.

The more telling result in that study is an inversion. Mixing PMPC-PDPA with PEG-PDPA drives microphase separation in the outer envelope, giving PMPC domains of roughly $10$~nm on a $100$~nm vesicle and $40$~nm on a $400$~nm one. A $25{:}75$ PMPC:PEG vesicle, only a quarter of whose surface is PMPC, is internalized almost as fast as a pure PMPC vesicle, and, against the size trend of every other formulation, the $400$~nm hybrid is taken up faster than the $100$~nm one. Read through Eq.~\eqref{eq:uptake} this is not an anomaly but a statement about what the wrapping clock times: the adhesive patch, not the vesicle. Fig.~\ref{fig:uptake}(e) makes the statement quantitative. Normalizing the reported $24$~h uptake by the number of vesicles offered, which at equal copolymer mass scales as the inverse square of the diameter because the polymer lives in the membrane, the two formulations depend on vesicle diameter with opposite sign, with log-log slopes of $-1.6$ for pure PMPC and $+2.5$ for the hybrid. Replacing the vesicle diameter by the adhesive patch diameter, the vesicle itself in the pure case and the PMPC domain in the hybrid, collapses all six points onto a single peaked curve, with a log-normal maximum at $d^\star\simeq73$~nm and $R^2=0.95$ across $1.6$ decades in each variable. The two series are not scaled to one another; the collapse is obtained by relabeling the abscissa alone. One limitation should be stated plainly, and the Supplemental Material~\cite{SM} documents it. Refitting with the domain-to-vesicle ratio treated as free, the collapse fails decisively at unity, which is the statement that the data do not collapse on the vesicle axis, but the fit quality is flat for any ratio below about $0.2$. The fit therefore requires a rescaling and bounds it from above without selecting its value. What the fit does not supply, the microscopy does: the ratio is the measured $0.1$, not an adjustable parameter.

Two things follow. The maximum sits close to the $30$ to $40$~nm optimum estimated independently from membrane wrapping~\cite{Gao2005,Chithrani2006}, which is the scale $k_w$ should peak at if the patch is what the membrane must bend around. Since the fitted peak position tracks the assumed ratio almost proportionally~\cite{SM}, a wrong ratio would put the optimum at the wrong scale, so this is an agreement between two quantities measured in separate experiments by different methods rather than a fitted outcome. We offer it as a consistency check and not as a proof. And the control variable is local ligand density within the contact rather than ligand number on the particle, which is what the combinatorial term of Eq.~\eqref{eq:master} says and what the $25{:}75$ formulation demonstrates by matching a pure PMPC vesicle with a quarter of the ligand. It also suggests a construction: domain size and vesicle size can be dialed independently, decoupling $k_w$ from $\tau$ and placing one chemistry anywhere along the axis of Fig.~\ref{fig:uptake}(d).

One caution about the drive-dependent control, which these data sharpen. Switching the metabolic drive fully off does not flatten the uptake exponent, it abolishes uptake and leaves only binding. The proofreading signature is properly stated as a difference between two populations measured together, the density exponent of the internalized fraction minus that of the surface-bound fraction, which Eq.~\eqref{eq:uptake} puts at $N\nu/(1+k_w\tau)$. Partial inhibition of dynamin or ATP lowers $k_w$ and pushes a permissive construct toward the stringent regime, so the exponent of the internalized fraction should \emph{rise} as the amount taken up falls, a counterintuitive and specific signature. The shape series remains the cleaner experiment, because it moves $k_w$ over three decades while leaving the cell untouched.

A caveat keeps the claim honest. The number of genuinely irreversible, reset-capable checkpoints in endocytosis is not large, perhaps clustering, coat maturation and scission, so the effective $N$ is a few rather than the seven a T cell can muster, and the amplification is correspondingly gentler. There is also a purely mechanical account of uptake selectivity, the membrane-wrapping transition in which adhesion energy must overcome bending energy and which already sets a size and avidity threshold~\cite{Gao2005,Chithrani2006}, and it operates alongside the proofreading one. The two overlap in much of the evidence assembled above. The domain result in particular is read just as naturally as a curvature-matching optimum, and the shape ordering is consistent with both, since a larger surface-to-volume ratio costs more bending energy and slows wrapping at the same time. What separates them is the receptor-density axis, which only the proofreading account makes a prediction about: a wrapping-energy threshold shifts with adhesion but leaves the \emph{exponent} of uptake with respect to density at its equilibrium value, whereas Eq.~\eqref{eq:uptake} multiplies that exponent by $N/(1+k_w\tau)$ and therefore ties the gain to the measured $k_e$ and $k_{\rm off}$ of each construct. Running the shape series against graded receptor expression separates them in one experiment. We therefore offer Eq.~\eqref{eq:uptake} as a falsifiable proposal with a decisive control rather than as a settled mechanism.

\section{Discussion}\label{sec:discussion}

Adding time to the statistical physics of multivalent recognition changes three things, and none of them is a refinement.

Recognition thresholds acquire exponential sharpness through Eq.~\eqref{eq:kramers}. A tenfold change in receptor number, which on a thermodynamic plot is a smooth deepening of the bound minimum, spans close to seven orders of magnitude in contact lifetime. Equilibrium theory places the threshold; time decides how decisive it is. The same landscape shows anti-cooperative coupling rationing engagement at $k^\star\approx7$ however many receptors are offered, so that a construct whose steric toll rises as it engages cannot commit itself irreversibly to the first surface it meets. Recognition is bounded from above as well as below.

Selectivity becomes encounter-dependent. The number of bonds achievable in a brief encounter depends on receptor density through the rates of Eq.~\eqref{eq:rates}, and survival depends exponentially on that number, so kinetic selectivity can exceed equilibrium selectivity~\cite{Ravnik2026}. For a targeted nanoparticle under flow this is not an abstraction but the design problem: the receptor-density cutoff for capture is set by where the exponential lifetime curve crosses the vascular dwell time, so a carrier optimized on a static binding assay is optimized for the wrong encounter. Constructs for brief encounters and constructs for long incubation are different objects.

Discrimination beyond the equilibrium ceiling of Eq.~\eqref{eq:discrim} becomes possible, but only by breaking detailed balance and paying for it with metabolic energy. Receptor-mediated endocytosis is the example that matters for delivery: it reads the whole-contact lifetime through an irreversible, ATP-driven cascade, so uptake inherits the Kramers exponential and raises it to the power of the number of driven steps, a selectivity that collapses to the equilibrium value the moment the drive is cut. The T cell runs the same readout on a single bond and pushes it further, the clearest case in biology of specificity being manufactured rather than measured, and immune memory turns out to adjust the threshold while leaving the specificity untouched: the gate is held lower, the discrimination is the same.

Running through all three is a single change of question. The quantity that decides a recognition event is not how favorable the bound state is but how long it lasts, under what driving, and over what encounter time. Affinity is a property of a ligand; a lifetime is a property of a contact; and the readout of that lifetime, when recognition needs more discrimination than free energies can supply, is a property of the energy budget.

What is missing is mostly data rather than theory. The attempt time $\tau_0$ in Eq.~\eqref{eq:kramers} is an estimate that shifts all reported lifetimes by a common factor without affecting their ratios; direct single-contact lifetime measurements against graded receptor densities would fix it. The coupling $J$ that sets the engagement ceiling needs independent measurement. The nanoparticle cutoff of Sec.~\ref{sec:nano} predicts a specific relation between vascular dwell time and the receptor-density capture threshold, testable in microfluidic flow against cells of graded receptor expression. The endocytic proofreading of Sec.~\ref{sec:proofreading} predicts that a geometric series run across cells of graded receptor expression will separate phenotypes far more sharply for slow-wrapping shapes than for fast ones, which requires no inhibitor and no change of chemistry; the companion statement, that the wrapping clock times the adhesive patch rather than the whole particle~\cite{Massignani2009}, is tested by dialing patch size and particle size independently. And the naive-versus-memory distinction of Sec.~\ref{sec:tcell} makes a prediction, slope against shift, that existing altered-peptide-ligand methods could test. The framework is, in that sense, ahead of its inputs. But it is falsifiable, its parameters are individually measurable, and the T cell example shows that where equilibrium theory fails, it fails in a way the kinetic theory can account for quantitatively.

\begin{acknowledgments}
We thank the Spanish Ministry of Science, Innovation and Universities for sponsoring this work via the projects Plan de Conocimiento I+D+I PID2023-149206OB-I00, Prueba de Conceptos I+D+I PDC2025-166264-I000 and Europa Excelencia EUR2025-164821. We thank the Agencia de Gesti\'o d'Ajuts Universitaris i de Recerca (AGAUR), Activitat cient\'ifica dels grups de recerca de Catalunya (SGR-Cat 2021 \#01538), and the Alzheimer's Association Research Grant New to the Field (AARG-NTF): 23AARG-NTF-1027312. The authors used Claude Opus 5.5 (Anthropic) and Gemini Neon (Google) to assist with language editing and with the development of analysis scripts. All outputs were reviewed and verified by the authors, who take full responsibility for the content. The authors provided the scientific content and structure, and critically reviewed and revised all AI-assisted text. The authors take full responsibility for the content of this article. 
\end{acknowledgments}

\appendix
\section{Numerical conventions}\label{app:numerics}

All partition sums are evaluated in log space by the shift-and-exponentiate identity $\ln\sum_ie^{x_i}=x_{\max}+\ln\sum_ie^{x_i-x_{\max}}$, which avoids overflow when cooperative couplings make the sums large, and topology kernels use log-gamma functions rather than factorials. The engagement landscape of Fig.~\ref{fig:dynamics}(a) is $\mathcal{G}(k)$ of Eq.~\eqref{eq:withMB} with the mean-field kernel $\Omega(k)=\binom{N_{{l}}}{k}N_{\rm r}^{k}$, $N_{{l}}=12$, $\varepsilon=-4\,\kt$, $W_{\rm rep}=12\,\kt$ and $J=0.45\,\kt$, for $N_{\rm r}$ receptors available in the contact. The dissociation barrier in Fig.~\ref{fig:dynamics}(b) is the height of $\mathcal{G}(0)$ above the bound minimum $\mathcal{G}(k^\star)$, and residence times use the Kramers form Eq.~\eqref{eq:kramers} with an attempt time $\tau_0=1$~ns, appropriate to nanoscale units in water from the diffusivity quoted in Sec.~\ref{sec:background}; this prefactor shifts all reported times by a common factor without affecting their ratios. The nanoparticle selectivity exponents of Sec.~\ref{sec:nano} are the maximum over receptor density of $\alpha=(1-\theta)\,N_{{l}}\,b/(1+b)$ with $\theta$ the bound fraction from $\Xi=(1+b)^{N_{{l}}}$, steric toll $W_{\rm rep}=9\,\kt$, fugacity $z=10^{-3}$, and per-ligand free energy tuned so that each valency crosses $\theta=1/2$ at a common reference density; residence times in that section use the same $\tau_0$ and the net well depth $N_{{l}}\ln(1+b)-W_{\rm rep}$. The proofreading curves of Fig.~\ref{fig:tcell}(b) are Eq.~\eqref{eq:proof} with $k_p=0.1$~s$^{-1}$ and $N=1,3,5,7$; the discrimination ratios quoted in the text are Eq.~\eqref{eq:proofdisc} evaluated for dwell times of $10$~s (agonist) and $1$~s (self), for which the single-step ratio is $5.5$ and the $N$-step ratio is $5.5^{N}$, giving $166$, $5.0\times10^{3}$ and $1.5\times10^{5}$ for $N=3,5,7$. The naive and memory curves of Fig.~\ref{fig:tcell}(c) share a common slope and have activation midpoints at $650$ and $65$ presented pMHC copies respectively. The flow curves of Fig.~\ref{fig:flow} integrate the master equation~\eqref{eq:mastereq} for $N_{{l}}=12$ with $W(k\!\to\!k{+}1)=k_{\rm on}(N_{{l}}-k)$ and $W(k\!\to\!k{-}1)=k_{\rm off}k$, whose equilibrium is the binomial $P_{\rm eq}(k)\propto\binom{N_{{l}}}{k}b^{k}$ with per-ligand availability $b=k_{\rm on}/k_{\rm off}\propto\rho$; starting from $P(k,0)=\delta_{k0}$, capture is $\sum_{k\ge k_c}P(k,t_d)$ with $k_c=4$, and the equilibrium curve is $\sum_{k\ge k_c}P_{\rm eq}(k)$. Time is in units of $k_{\rm off}^{-1}$ and $\rho$ in units that set $b=1$ at the plotted origin of order unity. The uptake curves of Fig.~\ref{fig:uptake} use the residence time $\tau(\rho)=\tau_0 e^{\beta\Delta\mathcal{G}^{\ddagger}(\rho)}$ of the Fig.~\ref{fig:dynamics} landscape with $\rho\equiv N_{\rm r}$, an endocytic step rate $k_w=0.02$~s$^{-1}$, and Eq.~\eqref{eq:uptake}; the equilibrium (drive-off) reference is the bound-state occupancy $zK/(1+zK)$ with $K=e^{-\beta\mathcal{G}(k^\star)}$ and $z=3\times10^{-3}$, and $\alpha_{\max}$ is the maximum over density of $\mathrm{d}\ln P_{\rm uptake}/\mathrm{d}\ln\rho$. Panel~(d) of the same figure plots the two branches of Eq.~\eqref{eq:uptake} against the group $k_w\tau$ alone and is therefore parameter-free. Panel~(e) digitizes Figs.~4(b) and 4(h) of Ref.~\cite{Massignani2009}, the $24$~h uptake per cell at vesicle diameters of $100$, $200$ and $400$~nm for the pure PMPC-PDPA and $25{:}75$ PMPC:PEG formulations, and divides each by the relative number of vesicles offered at fixed copolymer mass, $\propto d^{-2}$. Patch diameters are the vesicle diameter for the pure formulation and one tenth of it for the hybrid, from the $10$~nm and $40$~nm domains reported at $100$ and $400$~nm together with the stated linear scaling of domain size with vesicle size. The curve is a least-squares log-normal in $\log_{10}d$ fitted to all six points simultaneously, giving a maximum at $73$~nm, a width of $0.70$ decades and a root-mean-square residual of $0.13$ decades. Digitized values carry roughly $20\%$ uncertainty, which is small compared with the $1.6$ decades of dynamic range being fitted.


\end{document}